\documentstyle[aps,prl]{revtex}

\begin{document}

\draft
\preprint{draft}
\twocolumn[\hsize\textwidth\columnwidth\hsize\csname@twocolumnfalse\endcsname

\title{Gliding dislocations in a driven vortex lattice}

\author{Stefan Scheidl$^1$ and Valerii M. Vinokur$^2$}

\address{$^1$Institut f\"ur Theoretische Physik, Universit\"at zu
  K\"oln, Z\"ulpicher Str. 77, D-50937 K\"oln, Germany\\ $^2$Materials
  Science Division, Argonne National Laboratory, Argonne, IL 60439}

\date{3 February 1997}

\maketitle

\begin{abstract}
  The dynamics of dislocations in a two-dimensional vortex lattice is
  studied in the presence of a pinning potential and a transport
  current. In a vortex lattice drifting with velocity $v$ a glide
  velocity $V_{\rm d}$ of the dislocation with respect to the vortex
  lattice is found to decay like $V_{\rm d} \sim v^{-4}$ for large
  drive. From this result the velocity for the crossover between a
  regime of coherent elastic motion and a regime of incoherent plastic
  motion of vortices is estimated.
\end{abstract}


\vskip1pc]
\narrowtext

Periodic structures subject to quenched disorder are a generic model
for a vast number of physical phenomena. The recent upraise of
interest was motivated by the discovery of high temperature
superconductors (HTS) that set a physical realization of the random
system with tunable parameters and which are easily accessible in
experiments \cite{review}. As the statistical mechanics of the {\it
  static} periodic media in the presence of disorder reaches its
mature phase, the physics of {\it non-equilibrium} processes proposes
a diversity of unresolved challenging questions. In the static case it
is established that quenched pinning destroys the crystalline order of
the vortex lattice on intermediate spatial scales
\cite{review,lark,feig}. The long-range order of the vortex lattice is
reduced to the quasi-long-range order of the ``Bragg glass''
\cite{N90,GD94}. It is important to note that in the static case and
for weak pinning, the topological structure of the lattice is
preserved in a certain range of the phase diagram \cite{top-order}.
 
One of the central issues of {\em non-equilibrium} statistical physics
of dirty periodic media is the very nature of driven steady states. It
was observed \cite{copp,j,KV94} that the depinning transition can be
accompanied by plastic deformations of the driven periodic structures
leading to the destruction of the moving solid. For driven vortex
lattices it was proposed \cite{KV94} that at larger drives a true
dynamic phase transition occurs from an incoherently moving
non-equilibrium state to a coherently moving solid phase. This
conclusion was later extended to driven charge density waves
\cite{BF95}. Such dynamic transitions have indeed been observed in
experiments \cite{dm-exp} and simulations \cite{KV94,Ryu+96}. Note,
that even in the limit of strong driving disorder severely affects the
lattice structure on large length scales (in dimensions $d \leq 3$),
which becomes strongly anisotropic and keeps glassy features
\cite{GL96,MSZ96,BMR96}.

A phenomenology of this dynamic melting is built on the notion that
the driven periodic medium experiences pinning as temporally
fluctuating distortion in the comoving frame. A convenient heuristic
tool to investigate the dynamic transition was introduced in Ref.
\cite{KV94}, where the effect of temporarily fluctuating disorder was
described as an additional thermal noise with an effective ``shaking
temperature''. This temperature increases when the current is reduced
and can lead to a ``dynamic melting'' of the lattice structure. While
the phenomenological approach enables to estimate the position of the
transition line on the phase diagram, the nature of the dynamic
melting remains an open question. A microscopic understanding of
dynamic melting requires the analysis of topological defects, which
become free at the transition. In this Letter we present a first step
in this direction focusing on the study the dynamics of {\em
  independent} dislocations in the driven two-dimensional vortex
lattice. Note that in a {\it homogeneously} driven vortex lattice
without pinning dislocations and vortices move with the same average
velocity. Inhomogeneous pinning of different parts of the vortex
lattice causes shear strains which generate a fluctuating force acting
on dislocations. A motion of dislocation relative to the vortices
arises from correlations of this force on {\it intermediate} scales.
We examine the glide motion of the dislocations in the driven vortex
lattice in the limit of large vortex drive, weak disorder, and well
below the melting temperature of the pure vortex lattice. We calculate
the dislocation velocity perturbatively and find that dislocations are
retarded with respect to the vortices. The velocity where dislocations
get pinned by disorder provides a non-equilibrium estimate for the
crossover between elastic and plastic vortex flow.

In order to derive the equation of motion for the dislocation we first
consider the dynamic response of the vortex lattice to disorder and
thermal fluctuations. We restrict our discussion for simplicity to the
two-dimensional vortex lattice. This lattice is supposed to move with
an average velocity ${\bf v}$ as a consequence of the Lorentz force
${\bf f}_{\rm L}$ generated by electric transport currents. We label
vortices by their perfect lattice position ${\bf R}$ in the comoving
frame. Their actual position in the laboratory frame is denoted by
${\bf r}$. Then a displacement can be defined as usual by ${\bf
  u}({\bf R},t)={\bf r} ({\bf R},t)-{\bf v}t$.

Neglecting nonlocal effects the elastic energy of the vortex lattice
reads
\begin{equation}
  {\cal H}_{\rm el}=\frac 12 \int \frac{d^2q}{(2\pi)^2} \left\{c_{11}
    q^2 |{\bf u}^L({\bf q})|^2 +c_{66} q^2 |{\bf u}^T({\bf q})|^2
  \right\},
\end{equation}
where ${\bf q}$ runs only over the first Brillouin zone and ${\bf
  u}^{L,T}$ are the longitudinal / transverse projections of the
displacement field ${\bf u}({\bf q},\omega) = \int dt \ ({\phi_0}/{B})
\sum_{\bf R} \ e^{i \omega t - i {\bf q} \cdot {\bf R}} \ {\bf u}({\bf
  R},t)$ and ${\phi_0}/{B}$ is the area per vortex.  The pinning
potential $V({\bf r})$ is supposed to have correlations $\overline{
  V({\bf r}) V({\bf r}')} = \Delta({\bf r}- {\bf r}')$. The pinning
energy of the lattice is ${\cal H}_{\rm pin}=\sum_{\bf R} V({\bf
  r}({\bf R}))$.

The motion of vortices follows the overdamped Langevin equation
\begin{equation}
\label{eq.mo}
\eta \dot {\bf r} ({\bf R},t) = - \frac{\partial}{\partial {\bf r}
  ({\bf R},t)}[{\cal H}_{\rm el}+{\cal H}_{\rm pin}] + {\bf f}_{\rm L} +
{\bbox \xi } ({\bf R},t)
\end{equation}
with the Bardeen-Stephen friction coefficient $\eta$ and a thermal
noise correlation $\langle \xi_\alpha({\bf R},t) \xi_\beta({\bf
  R}',t') \rangle = 2 \eta T \delta_{\alpha \beta} \delta_{{\bf
    R} {\bf R}'} \delta (t-t')$.

In the limit of large driving forces ${\bf f}_{\rm L}$ the response of
the vortices to the pinning potential vanishes. Then one has to
leading order ${\bf v}={\bf f}_{\rm L}/\eta$ and the external force
acting on the vortices is ${\bbox f} ({\bf R},t) := {\bbox \xi } ({\bf
  R},t)-{\bbox \nabla } V({\bf R+v}t)$.  Its correlations read
\begin{eqnarray}
\label{f.corr}
\overline
{\langle f_\alpha ({\bf q}, \omega)  f_\beta ({\bf q}', \omega') \rangle}
=\Xi_{\alpha \beta} ({\bf q}, \omega) 
\delta ({\bf q} + {\bf q}') \delta(\omega + \omega'),
\\
\Xi_{\alpha \beta}({\bf q}, \omega) = 2 \eta T \frac {\phi_0}{B}
\delta_{\alpha \beta}
+ \sum_{\bf Q} k_\alpha k_\beta  \Delta ({\bf k}) 
\delta (\omega+{\bf k} \cdot {\bf v}), \nonumber
\end{eqnarray}
where ${\bf k}:= {\bf Q}+{\bf q}$ with a reciprocal lattice vector
${\bf Q}$. In linear response \cite{SH73} the vortex displacements are
obtained from the forces ${\bf f}$ through the response function with
longitudinal part $\Gamma^L ({\bf q},\omega)=[-i\eta \omega +c_{11}
(\phi_0/B) q^2]^{-1}$ and transverse part $\Gamma^T ({\bf
  q},\omega)=[-i\eta \omega +c_{66} (\phi_0/B) q^2]^{-1}$. Typically
the longitudinal response is small compared to the transverse response
since $c_{11} \gg c_{66}$.

As one can immediately recognize from expression
(\ref{f.corr}),  pinning forces
are more relevant than thermal forces\cite{GL96}
on {\em large} length and time scales: thermal fluctuations
give a finite contribution to $\Xi_{\alpha \beta}({\bf q}=0,
\omega=0)$, whereas pinning forces give singular contributions for
${\bf Q}\neq 0$ with ${\bf Q} \cdot {\bf v}=0$. This effect is due to
the periodic structure of the vortex density and leads to the
destruction of transitional long-range order for all velocities ${\bf
  v}$ in dimensions $d\leq 3$ \cite{GL96,BMR96}. For any finite
driving the correct asymptotic behavior of the lattice roughness
cannot be obtained in a linear response approach. However,
as we show below, dislocation dynamics is governed by the elastic
response of the vortex lattice on intermediate scales where this approach is
sufficient.

The topological structure of the vortex lattice is expected to be
destroyed by dislocations and disclinations, at least for sufficiently
high temperature, strong disorder, and/or weak driving. Within an
elastic approach vortices are found to move like in channels
\cite{GL96,MSZ96} when the lattice is moving along one of its
principal directions.  The displacement field component having the strongest 
fluctuations is the component parallel to the velocity \cite{MSZ96,BMR96}.
This favours dislocations with Burgers vectors
parallel to the velocity.

The motion of such a dislocation is easiest in the glide-direction
parallel to the Burgers vector ${\bf b}$ with length of the lattice
spacing $a\approx \sqrt{\Phi_0/B}$. In this direction it can move by
slips of individual vortices. A motion of the dislocation
perpendicular to the Burgers vector (climb) is possible only in
combination with the creation of vacancies or interstitials. In other
terms, glide is controlled by the Peierls barrier which is much lower
than the energy barriers for dislocation creep. In what follows we 
thus can ignore dislocation climb. The dislocation motion is then
one-dimensional 
in the direction
parallel to the Burgers vector and to the velocity. Its position along
this direction in the {\em moving} vortex frame is denoted by $X(t)$,
and we can consider $Y(t)=0$ fixed.

According to the theory of dislocations\cite{theo_dis} internal
stresses 
\begin{equation}
\sigma_{\alpha \beta} = 2c_{66} \nabla_\alpha  u_\beta +
 (c_{11}-2c_{66}) \nabla_\gamma u_\gamma \delta_{\alpha \beta}
\end{equation}
of the lattice, due to thermal fluctuations and pinning, induce the
so-called Peach-Koehler (PK) force
\begin{equation}
K_\alpha=\varepsilon_{\alpha \beta}\sigma_{\beta \gamma} b_\gamma
\end{equation}
acting on the dislocation, where ${\bf \varepsilon }$ denotes the
antisymmetric tensor. In addition, the dislocation is
subject to a periodic potential $W(X)\approx -(W_{\rm p}/2) \cos(2 \pi
X/a)$ with amplitude $W_{\rm p} \approx 10^{-2} c_{66} a^2$ representing the
Peierls barrier for glide motion.  The resulting equation of motion for the
dislocation reads
\begin{equation}
\label{eqmo.disl}
  \eta_{\rm d} \dot{X}(t) = K_x(X(t),t)- \nabla_x W(X(t)) .
\end{equation}
Here the dislocation friction coefficient is taken equal to its low
velocity value $\eta_{\rm d} \approx (\eta/8 \pi) \ln \left(1/ n_{\rm
    d} a^2 \right)$ \cite{B69,K92} for an average dislocation density
$n_{\rm d}$. We assume that the vortex lattice is still ordered on
intermediate length scales, i.e. the density of dislocations is small
compared to the vortex density and $\eta_{\rm d} \gg \eta$.

The fluctuating PK force includes thermal and disorder induced parts.
The corresponding spatio-temporal correlator follows within linear
response straightforwardly from Eq.~(\ref{f.corr}) and has has a
rather complicated form. For the further analysis we single out the
most important contributions.

One contribution arises from the thermal noise ${\bbox \xi}$ acting on
the vortices and it is present even in the absence of disorder and a
driving force. It can be described as effective thermal noise
$\xi_{\rm d}(t)$ with correlations $\langle \xi_{\rm d}(t) \xi_{\rm
  d}(t') \rangle \approx 2 \eta_{\rm d} T \delta(t-t')$ experienced by
{\it dislocations}.

The other contribution comes from pinning forces generating 
fluctuating  displacements ${\bf u}({\bf R},t)$. As pointed out in
Ref. \cite{GL96},
the modes ${\bf Q}$ with ${\bf Q} \cdot {\bf v}=0$ of the force
correlator (\ref{f.corr}) generate displacements, which are stationary
in the laboratory frame, ${\bf u}^{(0)}({\bf R},t)={\bf u}^{(0)}({\bf
  R}+{\bf v}t)$. The associated contribution to the PK force is also
stationary in the laboratory frame and can be the only origin of a
pinning of dislocations by disorder. The correlations of these
contributions $K_x^{(0)}$ at $Y=0$ are
\begin{eqnarray}
\label{chi}
\overline {K_x^{(0)}(q_x) K_x^{(0)}(q_x')}& =& \chi(q_x) \delta(q_x+q_x'), \\
\chi(q_x) &\approx& \frac{16 a^2 c_{66}^2 \Delta_0 }{3 \pi a \eta^2 v^2} 
\left( \frac{\eta v |q_x|}{c_{66}^2} \right)^{1/2}. \nonumber
\end{eqnarray}
Here $\Delta_0:=\sum_{{\bf Q}: Q_x=0} Q_y^2 \Delta({\bf Q})$ and the
obtained formula is valid for $\eta v \gg c_{11} a$.  Although the
mode ${\bf Q}=0$ also gives a stationary contribution, it is
negligible as compared to the modes ${\bf Q}\neq 0$ with ${\bf Q}
\cdot {\bf v}=0$, because of $c_{11} \gg c_{66}$. We ignored the modes
${\bf Q} \cdot {\bf v} \neq 0$ which represent a shaking of the
dislocation due to the discreteness of the vortex lattice and are
expected only to renormalize the effective dislocation temperature.
This effect is negligible for large ${\bf v}$, and does not contribute
to the dislocation glide.

Having made the above approximations we rewrite the equation of motion 
(\ref{eqmo.disl}) determining the dislocation glide velocity
in a form
\begin{equation}
\label{eqmo.disl.2}
\eta_{\rm d} \dot{X} =K_x^{(0)}(X+vt) -
\frac {\pi W_{\rm p}}a \sin \frac{2 \pi X}{a}+\xi_{\rm d}. \nonumber
\end{equation}
We restrict ourselves to temperatures $T \ll W_{\rm p}$, i.e. well
below the melting temperature $T_{\rm KT} \approx (3 c_{66} a^2)/8
\pi$ of the pure lattice \cite{KT73}.  In this case the dislocation
moves via thermal activation from one minimum of the Peierls potential
to a neighboring one favored by the PK force. In the driven lattice
disorder biases this motion in the direction opposite to the vortex
drift.

For weak disorder $\Delta \ll T^2$ this effect is conveniently
calculated perturbatively in the dynamical functional formalism
\cite{MSR}, where the dislocation position $X$ and its response field
$\tilde{X}$ are assigned a statistical weight $\exp(-{\cal A})$ with a
dynamic action
\begin{eqnarray}
\label{def.act}
{\cal A}&=&{\cal A}_0+{\cal A}_1 , \\
{\cal A}_0&=& \int dt \{ \eta_{\rm d} T \tilde{X}^2 + i \tilde{X}
[\eta_{\rm d} \dot X+\nabla_X W(X)] \} \nonumber , \\
{\cal A}_1&=& \frac 12 \int dt \int dt' \tilde{X} \tilde{X}'
\chi(X-X'+v(t-t')) . \nonumber 
\end{eqnarray}
Here the disorder average is already performed and abbreviations $X
\equiv X(t)$, $X' \equiv X(t')$ etc. are used.

In the absence of disorder, dynamics is controlled by the free action
${\cal A}_0$. This action is minimized by instanton trajectories which
start from a minimum/maximum of $W$ at $t=-\infty$ and end at a
neighbored maximum/minimum at $t=\infty$:
\begin{mathletters}
\label{instanton}
\begin{eqnarray}
X_{\rm i}(t)&=& \sigma_2 \frac {a}{\pi} \arctan \left[ \exp \left(
    \sigma_1 \frac t{t_{\rm i}} \right) \right] , \\
\tilde{X}_{\rm i}(t)&=& -\frac{i}{2 \eta_{\rm d} T} [ \eta_{\rm d} \dot
  X_{\rm i}(t) + \nabla_X W(X_{\rm i})],
\end{eqnarray}
\end{mathletters}
where $t_{\rm i}=a^2 \eta_{\rm d}/2 \pi^2 W_{\rm p}$ is an instanton
time scale and $\sigma_1, \sigma_2=\pm1$ specifies the type. For
$\sigma_1=\pm1$ motion is up or down from a minimum to a maximum of
$W$ or vice versa.  For $\sigma_1 \sigma_2=\pm1$ motion is
forward/backward in $X$ direction. Both ``down''-instantons have
$\tilde{X}_{\rm i}^{\rm down}(t)=0$ and thus ${\cal A}_0^{\rm
  down}=0$. Both ``up''-instantons have
\begin{mathletters}
\label{up}
\begin{eqnarray}
  {\cal A}_0^{\rm up} &=& W_{\rm p}/T ,\\ 
\tilde{X}_{\rm i}^{\rm up}(t) &=& -
  \frac {i \sigma_2 a}{2 \pi t_{\rm i} T} \frac 1{\cosh (t/t_{\rm i})}.
\end{eqnarray}
\end{mathletters}
Eq.~(\ref{up}a) gives the usual statistical weight of thermally
activated instantons.

Now one can easily calculate the disorder corrections to the instanton
actions in perturbation theory. To evaluate ${\cal A}_1$ in the lowest order
in
$\Delta$ one has to
use the {\em unperturbed} instanton solutions and
recognizes immediately that ${\cal A}_1^{\rm down}=0$ since
$\tilde{X}^{\rm down}=0$. Near its saddle-point the dislocation has a
velocity $v_{\rm i}=v+\dot X_{\rm i}(0)$ relative to the disorder,
which is different for ``forward'' and ``backward'' instantons. In
terms of this velocity the action correction reads
\begin{eqnarray}
\label{A_1}
{\cal A}_1 &\approx& \frac 12 \int dt \int dt' 
\tilde{X} \tilde{X}'\chi(v_{\rm i} (t-t'))
\nonumber \\
&\approx&  - \frac{2 \Delta_0}{3 \pi^2 T^2}
\left( \frac{2 a^2 c_{66}}{\pi \eta v v_{\rm i} t_{\rm i}} \right)^{3/2}.
\end{eqnarray}

The resulting dislocation glide velocity $V_{\rm d}:=\langle \dot X
\rangle$ is obtained from the difference of activation rates for
forward and backward instantons, $V_{\rm d} \approx
(a/t_0)[\exp(-{\cal A}^{\rm fw, up})- \exp(-{\cal A}^{\rm bw, up})]$.
The attempt frequency of the dislocation $t_0\approx a^2 \eta_{\rm d}/
\pi W_{\rm p}$ is estimated according to Kramers rate theory
\cite{kramer}. In terms of the dislocation diffusion constant
$D_0=(a^2 /t_0) \exp (-{\cal A}_0)$ in the pure lattice we eventually
find a drift velocity
\begin{equation}
\label{v.glide}
V_{\rm d} \approx - \frac{D_0 \Delta_0}{v t_{\rm i} T^2}
\left(\frac{2 a^2 c_{66}}{\pi^3 \eta v^2 t_{\rm i}} \right)^{3/2}
\propto v^{-4}
\end{equation}
in the limit $v \to \infty$, where  ${\cal A}_1 \ll 1$.

Several properties of this resulting drift velocity are noteworthy: it
decays with the fourth power of velocity, which is much faster than
$V_{\rm d} \sim v^{-1}$ that one might expect on the basis of symmetry
arguments. The drift velocity is independent of $c_{11}$ for $c_{11}
\gg c_{66}$, where only shear modes of the lattice get excited. In the
considered temperature range $T \ll W_{\rm p}$ the drift velocity
arises from a small bias of instantonic hops. Since this process is
thermally activated the drift velocity increases monotonously with
temperature. The asymptotic form (\ref{v.glide}) holds only down to a
certain minimum temperature determined by the condition ${\cal A}_1
\ll 1$, which however is satisfied for large velocity.

From Eq.~(\ref{v.glide}) we can estimate the critical velocity where a
crossover from ``elastic'' to ``plastic'' flow takes place. In the
``elastic'' flow regime vortices flow essentially coherently.
Topological defects may be present due to thermal fluctuations.
Although we found that dislocations are always retarded with respect
to the vortices, this effect is small for high drift velocities $v \gg
|V_{\rm d}|$. On the other hand there is the ``plastic'' flow regime,
where dislocations are essentially pinned by impurities.  Such
dislocations are a source of rearrangements of neighborhoods between
vortices. This plasticity becomes significant for $1-|V_{\rm d}|/v \ll
1$. Although our derivation of Eq.~(\ref{v.glide}) is strictly valid
only for large velocities, we expect it to give a qualitatively
correct estimate for the velocity
\begin{equation}
\label{v_c}
v_{\rm c} \approx  \left[ \frac {D_0 \Delta_0}
{t_{\rm i} T^2} 
\left(\frac {2 a^2 c_{66}}{\pi^3 \eta t_{\rm i}} \right)^{3/2} \right]^{1/5}
\end{equation}
of the crossover between both regimes.

Although the instantaneous forces acting on the dislocation arise from
local stresses in the vortex lattice, the dislocation glide velocity
results from the correlations of these stresses on {\em intermediate}
scales $|X| \lesssim v t_{\rm i}$ for large drift velocities:
According to Eq. (\ref{up}b) the quantity $\tilde X_i(t)$ decays
exponentially on the instanton time scale $t_{\rm i}$. Thus the main
contributions to ${\cal A}_1$ in Eq. (\ref{def.act}) come from
$\chi(X)$ for $|X| \lesssim v t_{\rm i}$. This scale increases
$\propto v^1$, whereas the dynamic Larkin length, up to where the
perturbative calculation of $\chi$ is reliable, increases $\propto
v^3$ \cite{GL96,BMR96}. It is for this reason that the calculation of
$\chi$ in linear response approximation for the vortex lattice is
sufficient for the dislocation dynamics at large drive.

Relating our result $v_{\rm c} \propto \Delta_0^{1/5}$ with the
finding \cite{KV94} of a dynamic freezing velocity $v_{\rm f} \propto
\Delta_0$, one has to distinguish the case of weak and strong pinning.
For {\em strong} pinning $v_{\rm c}<v_{\rm f}$. Expression (\ref{v_c})
for $v_{\rm c}$ is then formally invalid since the lattice is no
longer ordered at these velocities. In this case we can describe only
the regime $v> v_{\rm f}$, where the vortex lattice preserves its
crystalline order on large scales and dislocations move with the
vortices. In the limit of {\em weak} pinning both velocities $v_{\rm
  f} < v_{\rm c}$ become small and the perturbative findings are no
longer quantitatively reliable. Nevertheless, the smallness of the
velocities indicates that the lattice is well ordered and dislocations
move essentially with the lattice at practically all drift velocities.
Qualitatively, our finding suggests a crossover between practically
elastic flow for $v>v _{\rm c}$ and strongly plastic flow for $v<v
_{\rm c}$. This crossover can be viewed as precursor of the pinning
transition.

To sum up, our result $V_{\rm d}\propto v^{-4}$ implies that the
motion of the vortex lattice in disorder is {\em always} accompanied
by plasticity, even if it is extremely weak at high drives. For weak
pinning we have estimated the characteristic velocity of the crossover
between ``elastic'' and ``plastic'' flow of the vortex lattice. We
expect these results to be valid also after the inclusion of
interactions between dislocations. For example a
dislocation-antidislocation pair will be retarded with respect to the
vortex lattice essentially as a whole, since the resulting glide
velocity is independent of the dislocation type. But even if this pair
remains bound it can create point defects during its motion.
Interactions between dislocations are clearly relevant for a study of
the qualitative nature of the dynamic melting transition \cite{KV94},
we will address this issue in a forthcoming publication.

The authors gratefully acknowledge discussions with G.T. Zim\'anyi.
This work was supported from Argonne National Laboratory through the
U.S.  Department of Energy, BES-Material Sciences, under contract No.
W-31-109-ENG-38 and by the NSF-Office of Science and Technology
Centers under contract No.  DMR91-20000 Science and Technology Center
for Superconductivity and by Deutsche Forschungsgemeinschaft project
SFB341.

\end{document}